\documentclass{moriond}

\def\Journal#1#2#3#4{{#1} {\bf #2}, #3 (#4)}

\def\PRL{\em Phys. Rev. Lett.}
\def\PRD{{\em Phys. Rev.} D}

\def\mco{\multicolumn}

\def\ra{\rightarrow}

\def\ko{K^0}

\def\be{\begin{equation}}
\def\ee{\end{equation}}
\def\bea{\begin{eqnarray}}
\def\eea{\end{eqnarray}}

\def\amuhadlo{a_\mu^\mathrm{Had, LO}}
\def\amu{a_\mu}

\newcommand{\Photo}{\includegraphics[height=35mm]{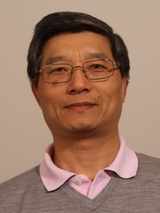}}

\begin{document}
\vspace*{4cm}
\title{DISPERSIVE APPROACH TO HVP FOR MUON G-2 }

\author{Z. ZHANG}

\address{Laboratoire de Physique des 2 infinis Irène Joliot-Curie – IJCLab, CNRS/Univ.\ Paris-Saclay\\
15 rue Georges Clémenceau, 91405 Orsay cedex, France}

\maketitle\abstracts{
The dispersive approaches to the hadronic vacuum polarisation contributions to the muon anomalous magnetic moment $a_\mu$ are described. The main issues are discussed followed by perspectives in the next years.}

\section{Introduction}

Vacuum polarisation (VP) originates from quantum fluctuations in the exchange of gauge bosons occurring in particle interactions. The muon anomalous magnetic moment $a_\mu$ is sensitive to such quantum fluctuations in all three sectors, QED, Weak and QCD, of the Standard Model.  
The measurement of $a_\mu$ has a long history and a strong interplay with the theoretical predictions~\cite{cern60}. The measurement is dominated by the recent Run-1 result of the Fermilab $g-2$ experiment~\cite{fermilab} and the earlier BNL measurement~\cite{bnl}. They are in good agreement and their average gives $116\,592\,061 (41)\times 10^{-11}$~\footnote{The units of $10^{-11}$ will be omitted for simplicity in the following.}. The comparison with the prediction of $116\,591\,810 (43)$ provided by the White Paper~\cite{wp} of the Muon $g-2$ Theory Initiative results in, however, a $4.2\sigma$ tension between the two. The theory prediction receives dominant contributions from QED with a negligible uncertainty. The hadronic vacuum polarisation (HVP) contribution, due to the fluctuations involving strongly interacting particles, includes the leading-order (LO), higher-order (HO) and light-by-light scattering components. The LO component, $\amuhadlo$, has the largest uncertainty of 40, to be compared with the total uncertainty of 43 for $\amu$. It is evaluated using the traditional dispersive approaches.

\section{How are the evaluations performed?}\label{sec:evaluation}

The HVP contributions are induced by fluctuations of quark-antiquark pairs. They can be computed at large energy scales in perturbative QCD but not at low scales due to the non-perturbative nature of QCD at large distance. Using unitarity and analyticity, the imaginary part of the two-point correlation (or HVP) functions is connected to spectra of hadron production from $e^+e^-$ annihilation via a spin-one photon propagator by the dispersion relation~\cite{B-M}:
\begin{equation}
\displaystyle
\amuhadlo=\frac{1}{3}\left(\frac{\alpha}{\pi}\right)^2\int^\infty_{m^2_\pi}ds\frac{K(s)}{s}R^{(0)}(s)\,.
\end{equation}
The integration kernel~\cite{G-deR} $K(s)/s\sim s^{-2}$ strongly emphasises the low-energy part of $R(s)\equiv \sigma(e^+e^-\to \textrm{hadrons})/\sigma(e^+e^-\to \mu^+\mu^-)$.

\begin{figure}
\centering
\includegraphics[width=0.61\columnwidth]{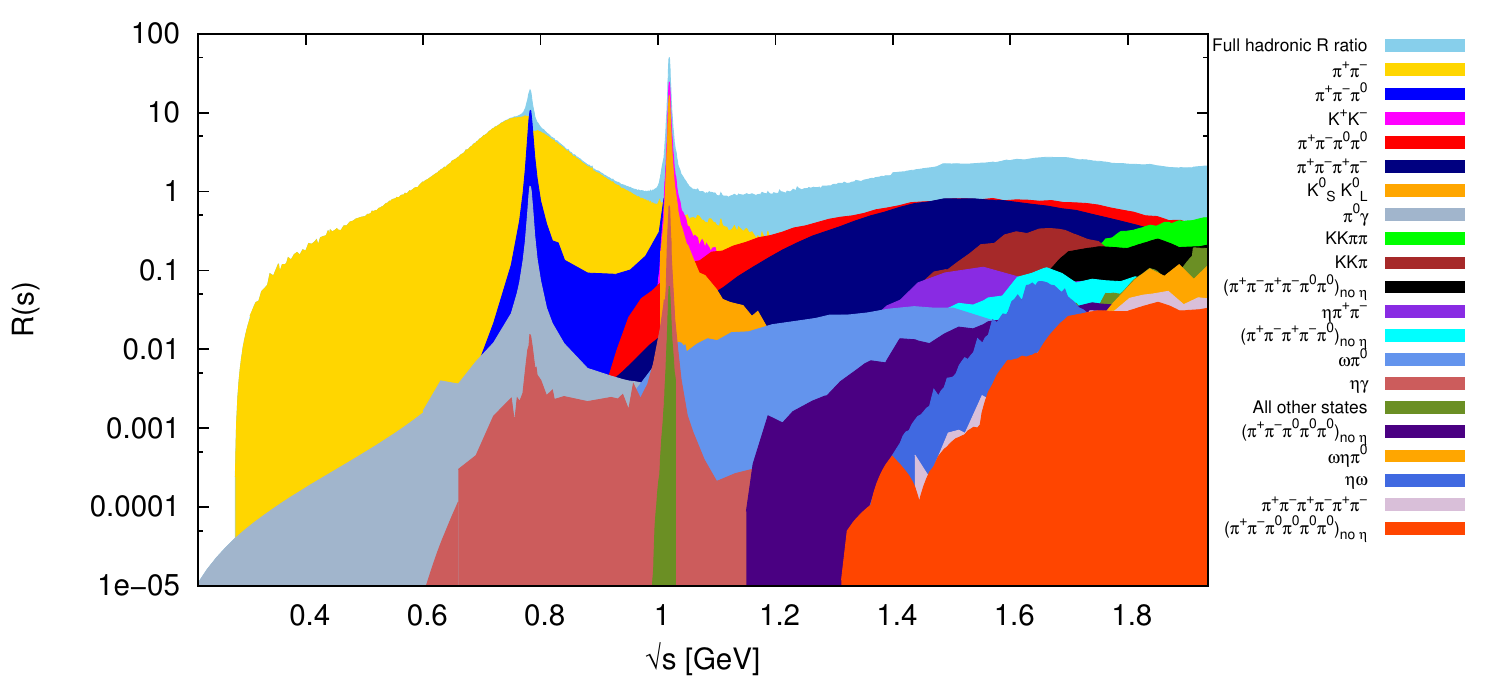}
\includegraphics[width=0.38\columnwidth]{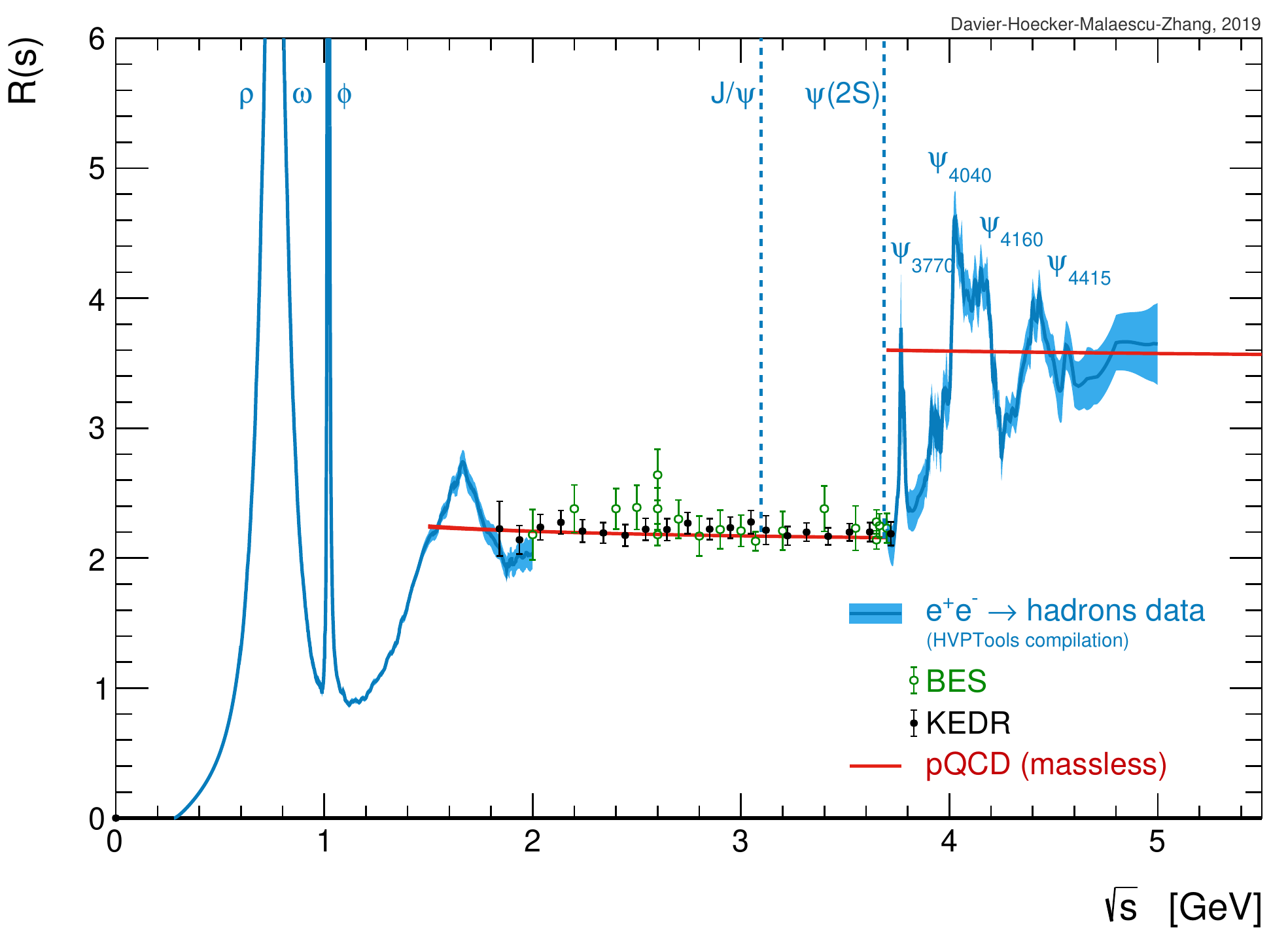}
\caption[]{Left: contributions to the total hadronic $R$-ratio from different final states (plot taken from Ref.~\cite{knt18}). Right: total hadronic $R$-ratio as a function of $\sqrt{s}$ (plot taken from Ref.~\cite{dhmz19}).}
\label{fig:channels}
\end{figure}

The HVP evaluation with the dispersive approaches is thus data-driven and the precision depends directly on the precision of the $e^+e^-\to \textrm{hadrons}$ cross-section data.
A large number of exclusive hadronic final states have been measured over a large energy range by many different experiments at $e^+e^-$ colliders in two general categories: 1) energy scan at different centre-of-mass energies $\sqrt{s}$, e.g.\ CMD2/3 and SND at VEPP-2M/2000; 2) varying energies $\sqrt{s^\prime}$ from a fixed value of $\sqrt{s}$ using the technique of radiative return with a hard photon emitted from the initial state radiation (ISR), e.g.\ KLOE at DAPHNE $\phi$-factory, BABAR at PEP-II B-factory, and BES\,III at BEPC\,II collider. The number of final states and their cross sections as a function of energy can be appreciated from Figure~\ref{fig:channels} (left). The dominant final state is $e^+e^-\to \pi^+\pi^-$ which contributes about 73\% to $\amuhadlo$ and 58\% to its uncertainty. 

Two groups, DHMZ and KNT, have performed since about 2 decades several complete evaluations of all exclusive final states below 1.8~GeV and 1.937~GeV, respectively. The two groups use different approaches for the data combination and uncertainty estimation. 

The combination and integration by DHMZ uses a dedicated package HVPTools deployed first in Ref.~\cite{hvptools}. It transforms the bare cross-section data and associated statistical and systematic covariance matrices into fine-grained bins, taking into account the correlations between the measurements of one experiment, as well as inter-experiment and inter-final-state correlations. The bin size is typically 1~MeV and smaller for the $\omega$ and $\phi$ resonances. The interpolation between 
adjacent measurements of a given experiment uses second-order polynomials, to avoid potential bias when using the linear trapezoidal interpolation. The averaging of the interpolated measurements from different experiments is performed using pseudo-data generated from the original measurements within wider averaging region so that all locally available experiments contribute. The uncertainty of the average is inflated if the local $\chi^2$/dof is greater than 1 following the PDG procedure. For the dominant $\pi^+\pi^-$ final state, the $\amuhadlo$ evaluation at energies below 0.6~GeV uses constraints from unitarity and analyticity. The evaluation of the total hadronic $R$-ratio for energies up to 5~GeV is shown in Figure~\ref{fig:channels} (right).

The data combination by KNT for a given final state is performed using an iterative procedure by minimising a $\chi^2$ function comparing the linear-model interpolant cross section values of clusters against direct measurements~\cite{knt18}. The cluster size depends on local data density and the correlation of measurement uncertainties is taken into account in the fit. The fit gives the mean $R$ values, a covariance uncertainty matrix, and local $\chi^2_\mathrm{min}$/dof for each cluster, with dof being the number of degree of freedom within a cluster. The quantity $\amuhadlo$ is derived using trapezoidal integration between clusters. The uncertainty of KNT's recent evaluations is also inflated using $\chi^2_\mathrm{min}$/dof.

The HVP prediction in the White Paper was obtained in a conservative merging procedure using the evaluations of these two groups~\cite{knt19,dhmz19} together with a partial evaluation in the $\pi^+\pi^-$ and $\pi^+\pi^-\pi^0$ final states~\cite{Colangelo:2018mtw,Hoferichter:2019mqg} with constraints from unitarity and analyticity. The procedure accounts for tensions among the measurements, for differences in methodologies in the combination of experimental inputs, and for correlation between systematic errors. 

\section{What are the issues?}

The main issue concerns the discrepancy between two most precise measurements from BABAR and KLOE in the dominant $\pi^+\pi^-$ final state. The DHMZ group studied the discrepancy by removing either of them from the evaluation and found a difference of 27.6 which is included in the channel-specific systematic uncertainty to this final state shown in Table~\ref{tab:KNT-DHMZ} (adapted from Table~5 in Ref.~\cite{wp}). Among the selected final states shown in the table, the difference between the two evaluations for some of the channels is larger than the quoted uncertainties from one evaluation reflecting again their different methodologies. 

Unfortunately, the $\pi^+\pi^-$ is not the only problematic case. Differences between different measurements are also observed in other final states, such as $K^+K^-$ around the $\phi$ resonance peak~\cite{dhmz19}. It should be noted that it is difficult to determine kaon detection
efficiency with precision for CMD-2/3 and SND and easier for BABAR with large ISR photon energies.

\begin{table}[htb]
\small
\caption{Selected exclusive-mode contributions to $\amuhadlo$ from DHMZ19 and KNT19, for the energy range $\leq 1.8$~GeV. Where three (or more) uncertainties are given for DHMZ19, the first is statistical, the second channel-specific systematic, and the third common systematic, which is correlated with at least one other channel.} 
		\label{tab:KNT-DHMZ}
		\vspace{0.2cm}
			\centering
	\begin{tabular}{|c| r| r| r|}
	\hline
	  & DHMZ19 & KNT19 & Difference\\\hline
	 $\pi^+\pi^-$ & $5078.5(8.3)(32.3)(5.5)$ & $5042.3(19.0)$ & $36.2$\\
	 $\pi^+\pi^-\pi^0$ & $462.1(4.0)(11.0)(8.6)$ & $466.3(9.4)$ & $-4.2$\\
	 $\pi^+\pi^-\pi^+\pi^-$ & $136.8(0.3)(2.7)(1.4)$ & $139.9(1.9)$ & $-3.1$\\
	 $\pi^+\pi^-\pi^0\pi^0$ & $180.3(0.6)(4.8)(2.6)$ & $181.5(7.4)$ & $-1.2$\\
	 $K^+K^-$ & $230.8(2.0)(3.3)(2.1)$ & $230.0(2.2)$ & $0.8$\\
	 $K_SK_L$ & $128.2(0.6)(1.8)(1.5)$ & $130.4(1.9)$ & $-2.2$\\
	 $\pi^0\gamma$ & $44.1(0.6)(0.4)(0.7)$ & $45.8(1.0)$ & $-1.7$\\\hline
	 Sum of the above  & $6260.8(9.5)(34.8)(14.7)$ & $6236.2(22.7)$ & $24.6$\\\hline
	 $[1.8,3.7]$~GeV (without $c \bar c$) & $334.5(7.1)$ & $344.5(5.6)$ & $-10.0$\\
	 $J/\psi$, $\psi(2S)$ & $77.6(1.2)$ & $78.4(1.9)$ & $-0.8$ \\
	 $[3.7,\infty)$~GeV & $171.5(3.1)$ & $169.5(1.9)$ & $2.0$\\ \hline
	 Total $\amuhadlo$ & $6940(10)(35)(16)(1)_\psi(7)_\mathrm{DV+QCD}$ & $6928(24)$ & $12$\\ \hline
	\end{tabular}
\end{table}

\section{Perspectives}

The Fermilab Run-1 result has an uncertainty of 54, which is statistics dominated. The uncertainty is expected to be improved by a factor of 2 with the subsequent runs in the next years and eventually by a factor of 4 with its final measurement. The uncertainty of the HVP prediction has to be reduced concurrently.

The most important and urgent task is to understand the discrepancy between BABAR and KLOE. To resolve the discrepancy, new ideas, improved event generators, and mostly new measurements with higher precision and using a blind approach are needed. 

Since the release of the White Paper, a new measurement of the $\pi^+\pi^-$ final state between 525 and 883~MeV by SND has been published~\cite{snd21}. Above $\sim 750$~MeV, the measurement, though with a limited systematic uncertainty of 0.8\% for $\sqrt{s}>600$~MeV, seems to be in better agreement with BABAR than with KLOE. There are also two new measurements for the $\pi^+\pi^-\pi^0$ final state by SND in the energy range of 1.15$-$2.0~GeV~\cite{snd20-3pi} and by BABAR in the energy range of 0.62$-$3.5~GeV~\cite{babar21-3pi}. The latter measurement is measured with a systematic uncertainty of 1.3\% in the energy regions of the $\omega$ and $\phi$ resonances. This final state has the second largest cross section after $\pi^+\pi^-$ below 1~GeV which contributes currently with a precision of 3\% to $\amuhadlo$. The BES\,III has published a new $R$ measurement at 14 energy points between 2.2324 and 3.6710~GeV with an accuracy of better than 2.6\% below 3.1~GeV and 3\% above~\cite{bes22-r}. The measured cross section is, however, higher than the corresponding KEDR result~\cite{kedr} and perturbative prediction, in particular in the energy regions 3.4$-$3.6~GeV by 1.9 and 2.7 standard deviations. There are also attempts to improve the uncertainty for the $\pi^0\gamma$ and $K\bar{K}$ final states~~\cite{pi0y,kkbar}  using constraints from analyticity, unitarity, and crossing symmetry as well as low-energy theorems, though they are not the dominant final states contributing to $\amuhadlo$.

One surprise after the publication of the White Paper is the Lattice prediction of $\amuhadlo$ by BMW~\cite{bmw20} which has dramatically improved the previous Lattice predictions by many factors to reach a precision of 0.8\% and has a value lying in between the direct measurements and the dispersive prediction with a deviation of $1.5\sigma$ and $2.1\sigma$, respectively. If this result will be confirmed with comparable precision by the other Lattice groups, the discrepancy between the dispersive and Lattice predictions must be understood.

\section*{Acknowledgments}

I am grateful for the fruitful collaboration with my colleagues
and friends M.~Davier, A.~Hoecker, and B.~Malaescu.

\end{document}